# 5G Radio Access Network Architecture for Terrestrial Broadcast Services

Mikko Säily, Carlos Barjau Estevan, Jordi Joan Gimenez, Fasil Tesema, Wei Guo, David Gomez-Barquero and De Mi

*Abstract* — The 3rd Generation Partnership Project (3GPP) has defined based on the Long Term Evolution (LTE) enhanced Multicast Broadcast Multimedia Service (eMBMS) a set of new features to support the distribution of Terrestrial Broadcast services in Release 14. On the other hand, a new 5th Generation (5G) system architecture and radio access technology, 5G New Radio (NR), are being standardised from Release 15 onwards, which so far have only focused on unicast connectivity. This may change in Release 17 given a new Work Item set to specify basic Radio Access Network (RAN) functionalities for the provision of multicast/broadcast communications for NR. This work initially excludes some of the functionalities originally supported for Terrestrial Broadcast services under LTE e.g. free to air, receive-only mode, large-area single frequency networks, etc. This paper proposes an enhanced Next Generation RAN architecture based on 3GPP Release 15 with a series of architectural and functional enhancements, to support an efficient, flexible and dynamic selection between unicast and multicast/broadcast transmission modes and also the delivery of Terrestrial Broadcast services. The paper elaborates on the Cloud-RAN based architecture and proposes new concepts such as the RAN Broadcast/Multicast Areas that allows a more flexible deployment in comparison to eMBMS. High-level assessment methodologies including complexity analysis and inspection are used to evaluate the feasibility of the proposed architecture design and compare it with the 3GPP architectural requirements.

*Index Terms* — 5G, architecture, broadcast, multicast, point-to-point, point-to-multipoint, radio access network, single frequency network, signal synchronisation.

## I. INTRODUCTION

THE 3rd Generation Partnership Project (3GPP) finalised the first set of 5th Generation (5G) specifications for Release 15 (Rel-15) in December 2018. This defines a new Radio Access Technology (RAT) known as New Radio (NR), the Next Generation Radio Access Network (NG-RAN) and the 5G Core Network which embrace several design principles such as: (i) forward compatibility with future releases; (ii) control-user plane separation (CUPS); (iii) lean and cloud-native system design. Rel-15 and Rel-16 only cover unicast, or Point-to-Point (PTP), transmissions. However, benefits of multicast and broadcast, or Point-to-Multipoint (PTM), have been already assessed as beneficial for some 5G use cases [1], [2].

The support of PTM communications is not new in 3GPP. Mobile broadcast as a service is already included in Long Term Evolution (LTE) as per the enhanced Multicast/Broadcast Multimedia Service (eMBMS). The set of specifications have been updated to support new services such as public safety, Internet of Things (IoT) or Vehicle-to-Everything (V2X) [3]. Its most recent update comes in Rel-14 [4] and Rel-16 [5], [6] in order to support the 5G requirements for broadcast, and in particular for the provision of Terrestrial Broadcast services. This has implied severe changes at the air-interface to implement larger Single Frequency Networks (SFNs) or the introduction of carriers with dedicated broadcast content. The architecture relies on the existing for eMBMS with the introduction of the so-called receive-only mode for Subscriber Identification Module (SIM)-free operation even without uplink, or a new xMB interface between the eMBMS system and service providers. So far, eMBMS has proven limited success among mobile network operators due to a demanding implementation both at the network architecture and user equipment.

Under the 5G System (5GS) and NG-RAN architectures, basic support for multicast/broadcast is expected to be introduced in Rel-17. This includes Multicast/Broadcast support at 5G Core Network [7] and NR-based Mixed Mode transmissions at RAN [8]. The support of Terrestrial Broadcast services is not in the scope of these specifications. However, a generic architectural solution that could allocate requirements from different domains would be beneficial to increase deployment opportunities.

One of the 3GPP system requirements for 5G is a flexible broadcast/multicast service for three types of devices, be they enhanced Mobile Broadband (eMBB), Ultra Reliable and Low

This work was supported in part by the European Commission under the 5GPPP project 5G-Xcast (H2020-ICT-2016-2 call, grant number 761498). The views expressed in this contribution are those of the authors and do not necessarily represent the project. Part of the material in this paper was presented at the IEEE International Symposium on Broadband Multimedia Systems and Broadcasting (BMSB) 2019 [20].

M. Säily is with the Standardization and Research Lab, Network & Architecture Group, Nokia Bell Labs, Espoo, 02610 Finland (e-mail: mikko.saily@nokia-bell-labs.com).

C. Barjau Estevan and D. Gomez-Barquero are with the Institute of Telecommunications and Multimedia Applications (iTEAM), Universitat Politecnica de Valencia, 46022 Valencia, Spain (e-mail: {carbare1, dagobar}@iteam.upv.es).

J. J. Gimenez is with the Future Networks Department, Institut für RundfunktTechnik GmbH, 80939 Munich, Germany (e-mail: jordi.gimenez@irt.de).

F. Tesema is with Nomor Research GmbH, Munich, 81541 Germany (e-mail: tesema@nomor.de).

W. Guo is with Samsung R&D Institute UK, Staines-upon-Thames TW18 4QE, U.K. (e-mail: wei6.guo@samsung.com).

D. Mi is with the 5G Innovation Centre (5GIC), Institute for Communication Systems, University of Surrey, Guildford GU2 7XH, U.K. (e-mail: d.mi@surrey.ac.uk).

Latency Communications (URLLC) and massive Machine Type Communications (mMTC) [9]. 5G should be envisioned as a system of systems, where the Core and Transport Network alongside the RAN must accommodate a plethora of different services, with stringent requirements, ranging from several gigabits/sec (think of Augmented/Virtual reality - AR/VR) to low kilobits/sec throughput (think of mMTC), latencies ranging from 1 millisecond (e.g., industrial IoT) up to several seconds (e.g., best-effort data delivery), mobility support to moving devices, from static equipment (e.g., roof-top antennas) up to 500 kilometre per hour (e.g., V2X) and support for millions of users per square kilometre (e.g., massive IoT) [10]-[12].

The 5G Infrastructure Public Private Partnership (5G-PPP) project 5G-Xcast has developed a holistic implementation of 5G PTM systems [13], covering Core Network [14]-[16] and RAN developments from air interface [17], [18] to architecture [19]-[21] and protocols [21], [22]. This can facilitate the fulfilment of requirements from different applications [23], including traditional Terrestrial Broadcast deployments, which are the scope of this work. The main contributions of this work are listed as follows:

- We develop a dynamic RAN Broadcast/Multicast Areas mechanism that allows the delivery of multicast/broadcast services wherever needed without the fixed deployment on top of the existing RAN.
- Our proposed RAN architecture design supports multi-cell and SFN transmissions using NG-RAN based synchronization method fulfilling Quality of Service (QoS) targets defined for traffic flows.
- We introduce a new RAN interface design to support Terrestrial Broadcast and multicast with the minimal impact on the current 5G system.
- We propose a detailed procedure and deployment strategy for our architecture design, together with high-level evaluations including complexity analysis and inspection, providing insightful and practical instructions on the feasibility of our design.

The rest of this paper is structured as follows. First, the RAN architecture evolution to 3GPP Rel-15 is discussed in Section II. Next, it details in Section III our design on the new RAN architecture for the Terrestrial Broadcast. Then, deployments and procedures of the proposed architecture are presented in Section IV. Section V provides a complexity analysis of our architecture design for both Terrestrial Broadcast and multicast scenarios. Finally, Section VI concludes the key findings and potential ways forward.

## II. PRELIMINARIES ON 5G NR RAN ARCHITECTURE IN 3GPP REL-15

The key architectural element in RAN design in 3GPP Rel-15 specifications is to extend the distributed base station architecture towards flexible Cloud-RAN based protocol functionality where the computing hardware pools are used to handle the higher layer processing of user plane data traffic and control plane signalling. The protocol functionality split of NR base station, namely Next generation NodeB (gNB), between

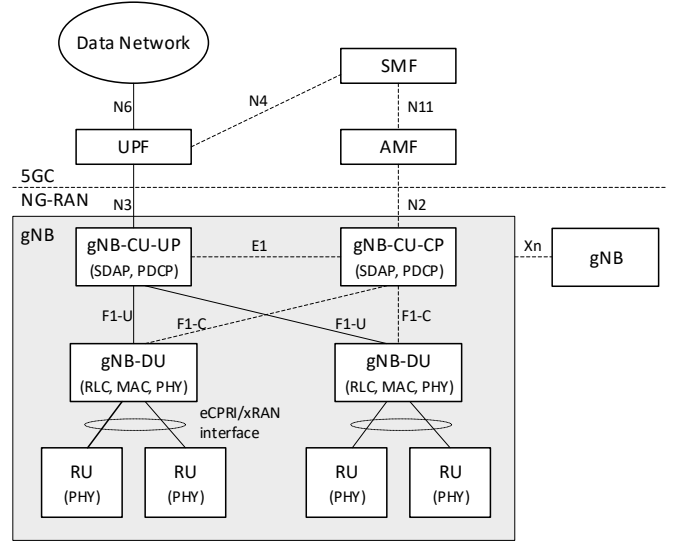

Fig. 1. Rel-15 NG-RAN architecture with a CU-DU split deployment.

Central Units (CUs) and Distributed Units (DUs) in 5G architecture enables dynamic adaptation of QoS functions depending on the real-time radio conditions, user density and dynamically controlled geographical area.

As shown in Fig. 1, the gNB functions are split into CU and DU, where CU covers higher layer protocol functions of Service Data Adaptation Protocol (SDAP) and Packet Data Convergence Protocol (PDCP), and DU entails lower layer protocol functions of Radio Link Control (RLC), Medium Access Control (MAC) and Physical Layer (PHY). In a typical Cloud-RAN deployment, the CUs are placed in a computing hardware pool and thus form the cloud. The gNBs are inter-connected through an Xn interface.

The F1 interface provides control (F1-C) and user (F1-U) plane connectivity between the CU and DU, enabling deployments with C/U-plane separation. The E1 interface provides connectivity between the user plane CU-UP and control plane CU-CP, enabling deployments with C/U-plane separation on the CU level. The interface also provides separation between the radio network and transport network layers, enabling the exchange of User Equipment (UE) and non-UE associated information. When F1 is separated into F1-C and F1-U, consequently the Xn inter-connecting the gNBs is separated into Xn-C on the control plane and Xn-U on the user plane. A gNB-CU is further separated logically into gNB-CU-CP and gNB-CU-UP, with E1 Application Protocol (E1AP) providing the signalling service between them.

These architecture enhancements provide a significant opportunity to design an innovative RAN architecture for multicast and broadcast services.

## III. PROPOSED NEW RAN ARCHITECTURE DESIGN FOR TERRESTRIAL BROADCAST

### A. Design Principles

Our prior investigations [24] and [14] set some of the limitations found in eMBMS at the air-interface and system architecture, respectively. Even more, in [25] it collects a series of limitations for the provision of Terrestrial Broadcast services.

In order to overcome the overall limitations of eMBMS, we provide novel technical developments using an enhanced NG-RAN architecture based on 3GPP Rel-15, which primarily focuses on broadcast/multicast capabilities to address requirements from multiple verticals and is also able to be configured, in a more static fashion, to cover requirements focused on Terrestrial Broadcast services. The key architectural enhancement is leveraged on CU-DU split specifications as specified in 3GPP NG-RAN [26], and our main target is to provide a solution with a high commonality with unicast, minimizing the additional implementation footprint.

5G broadcast and multicast services should, in general, be available in dynamic areas where the number of users during popular events (e.g., in stadiums) can be high and the user distribution within the multicast area very likely changes over the time. In addition, seamless switching between PTP and PTM transmissions, dynamic adjustment of the RAN multicast area based on user distribution (from one cell to several synchronised cells), and efficient multiplexing with PTP transmissions in frequency and time domain should be provided. To this end, a concept of RAN Broadcast/Multicast Area (RBMA) is developed to allow delivery of PTM services wherever needed without eMBMS-type of static deployment on top of the existing RAN [19], [25]. RBMA mechanism takes account the user activity, user mobility, number of devices and their geographical distribution [19].

The RAN is aware of UE's interest to receive data from Internet Protocol (IP) multicast group. Dynamic RBMA with synchronization point in NG-RAN can support multitude of deployments from a single cell DU to multiple cells under several DUs, still controlled by a single CU. The proposed RAN architecture may also support a multi-cell transmission using NG-RAN based synchronization method, where synchronised DUs participate to multi-cell transmission using a single CU as a point for transmission coordination. This approach enables over the air transmission of synchronised multicast/broadcast traffic while fulfilling the QoS targets defined for the traffic flows.

*B. RAN Broadcast/Multicast Area for Terrestrial Broadcast*

In the context of Terrestrial Broadcast, the RBMA is configured according to pre-defined coverage requirements and agnostic to the QoS that UEs actually experience (either they have uplink capabilities or not) [19], [24], [25].

A slightly different approach is followed to address the requirement on large area coverage where the use of SFN modes are avoided when possible as this has a severe impact on the air-interface design. To the contrary, Terrestrial Broadcast infrastructure is usually heterogeneous and relies on local, regional or nationwide transmitter areas in SFN or in Multi-Frequency Network (MFN) with some degree of frequency reuse. The support of concurrent delivery of both unicast and PTM services to the users from the same cell, with efficient multiplexing with unicast transmissions is also taken into account. The design approach should also support fixed,

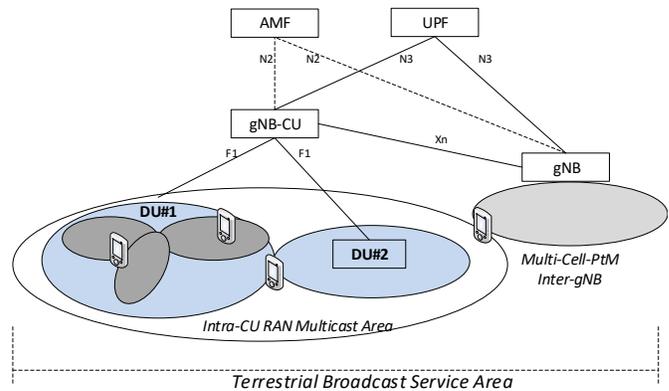

Fig. 2. Deployments of RBMA for Terrestrial Broadcast Service Area.

portable and mobile reception [24].

The RBMA for Terrestrial Broadcast Service Area, as illustrated in Fig. 2, is defined as the amount of time/frequency resources per transmitter area (either for a single transmitter, an MFN or SFN area) reserved for the potential transmission of Terrestrial Broadcast services. In order to adapt to a variety of deployments suitable for the delivery of Terrestrial Broadcast services, the RAN may be provided via Operation and Maintenance (O&M) with the list of cells that constitute a given RBMA with the following assumptions:

- Each single cell transmitter is considered as a constituent RBMA;
- A cluster of cells that constitute an SFN is regarded as a unique RBMA;
- A wide coverage area comprising a variety of topologies (e.g. mixture of single and SFN transmitter areas) is formed by means of multiple RBMA;
- One transmitter can be operating more than a single carrier, therefore, each cell in the list may be associated a given frequency (e.g., DL_EARFCN).

Each RBMA is identified by means of a RBMA Index (RBMA ID) which can be selected by the service provider via xMB interface [27], [28]. A 5G-Xcast Control plane network Function (XCF) is proposed to translate the RBMA ID to the actual identifiers of the gNBs [14], [19].

The amount of available resources per carrier might be different in each transmitter due to several circumstances such as the presence of other services, the use of carriers of different bandwidth or the needs of inter-transmitter scheduling (e.g. time/frequency reuse) to avoid interferences. Therefore, each RBMA shall be informed of the specific amount of time/frequency resources that need to be available for potential service scheduling via xMB. It is a design assumption in our RAN architecture that services to be transmitted in SFN will be scheduled over dedicated resources with an adequate numerology. The group of resources with different numerologies can be multiplexed by using different Carrier Bandwidth Parts (Frequency Division Multiplexing - FDM; multiplexed within a given OFDM symbol) or subframes / frames (via Time Division Multiplexing - TDM).

The delivery of each broadcast service, e.g. TV or radio, can be configured according to the RBMA ID where the service is meant to be delivered. Associated to each broadcast service, the Modulation and Coding Scheme (MCS) index that fulfils the robustness (coverage) and data rate requirements of the Service-Level Agreement (SLA) is indicated together with scheduling information in terms of required time/frequency resources for the given data rate (e.g. initial and final Physical Resource Block - PRB).

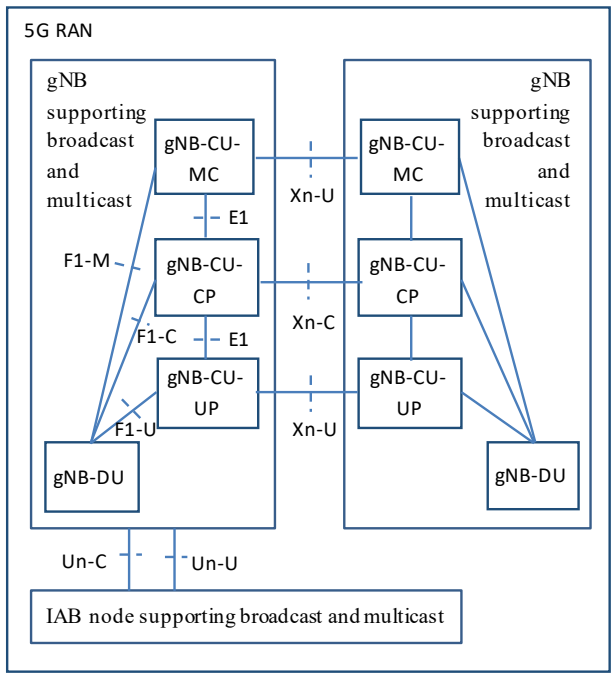

Fig. 3. 5G RAN internal architecture and interfaces.

An admission control procedure will determine the allocation of a new broadcast service according to the amount of available resources in the carrier for the allocation of Terrestrial Broadcast service (as indicated per RBMA) and the amount of required resources per service.

*C. Synchronised Content Delivery*

To fulfil the SFN requirements, the 5G-Xcast RAN incorporates two main functionalities, one involving the control plane residing in the gNB-CU-C and the other related to the user plane inside the gNB-CU-MC (as shown in Fig. 3): The control plane part is the setup of the SFN area inside cellular networks, deciding the physical layer parameters such as modulation, code rate and scheduling to satisfy specific QoS. This decision is propagated using new signalling towards the relevant gNB-DU, which relays this to the relevant Remote Radio Heads (RRH). In addition, the gNB-CU can take into account existing unicast measurement reports to fine tune the physical layer parameters of the SFN transmission.

The second functionality is the constant encapsulation of the multicast data to provide Time-to-Air (TTA) information for the cells involved in the SFN transmission. A modified eMBMS synchronisation protocol (SYNC) based on [29] is used as the encapsulation protocol, but instead of manually setting the SYNC parameters between the eMBMS Core and the eNBs, the parameters are negotiated in the SFN setup process of the gNB-CU [20]. More specifically, the SFN parameter negotiation origins from the master gNB-CU that wants to activate a synchronised multicast transmission across many DUs and CUs. In this setup, relevant SYNC parameters like SYNC period and SYNC sequence are defined which are needed for the SYNC protocol. The revision of SYNC is called RAN-SYNC and is one of the main 5G-Xcast contributions [19], [20]. This approach enables fast and flexible network deployments and simplifies the network operating and maintenance process. In this case, the entity encapsulating the data resides inside the RAN, while in 4G eMBMS, SYNC is applied at the Broadcast Multicast Service Centre (BM-SC). Thus, the realization complexity to set up the network with the proposed RAN-SYNC can be lower than 4G eMBMS, as in eMBMS the operator must set up both Core and RAN but here the RAN can operate independently from the Core and its agnostic to the transport network used. Also note that we propose the use of SYNC across gNBs due to the fact that it allows the underlying gNB modules to reuse existing eMBMS technology (e.g., Multi-cell/multicast coordination entity - MCE) thus lowers the implementation costs.

The proposed RAN architecture does not include a dedicated network configuration entity, by which functionality would include the configuration of multi-cell transmission. Instead, the approach uses run-time configuration of the transmission parameters. In the multi-cell transmission, the transmitting gNB-DUs must be synchronised. The gNB-DUs exchange the information about their PHY synchronization/clock and system reference frame number, if this information is not readily available. The PHY synchronization and reference clock information could indicate a synchronization region such as Multicast Broadcast SFN (MBSFN) synchronization area Identity (Id) in eMBMS. The gNB-DUs can also determine whether they are synchronised to a common time reference (e.g. in Global Navigation Satellite System) and provide the PHY synchronization/clock information and system frame number as an offset to the common time reference. The latter approach does not require additional configuration between gNB-DUs.

*D. New RAN Interfaces for Broadcast*

ITU-T Recommendation I.112 [30] defines an interface as "the common boundary between two associated systems", and 3GPP follows the definition as in 3GPP TR 21.905 [31]. A network interface covers all protocol layers of significance for network elements at both sides of the interface. E.g. if the network elements are Layer 2 entities then the interface should be specified at Layer 2. In most cases, the interface specification goes all the way down to Layer 1, to be represented as a full protocol stack to enable the interconnection and even plug and play. RAN interfaces are categorised into external and internal interfaces. The external interfaces are those between the 5G RAN (named NG-RAN in 3GPP) and 5G Core Network (5GC), and those between the 5G RAN and the UE. The internal interfaces are those between 5G RAN nodes. 3GPP has been continuously working on the definition and standardization of those interfaces in the 5G system. The principles for 5G RAN interface design to support Terrestrial Broadcast are:

- To reuse as much as possible and to enhance current NG-RAN interfaces to support broadcast and multicast to keep the system interfaces as simple and as few as possible.

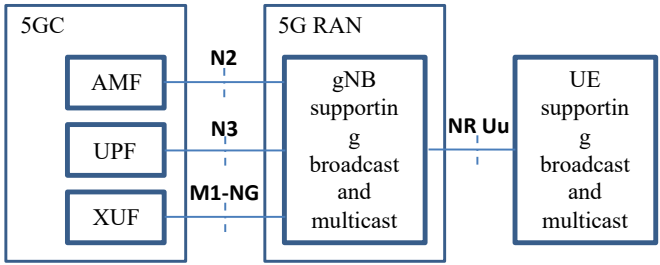

Fig. 4. Deployments of RBMA for Terrestrial Broadcast Service Area.

- To define new interfaces to support broadcast and multicast if it is necessary.

The network interfaces should allow easy interconnection of products from different vendors, and the possibility of forward compatibility for future evolution.

3GPP has defined the interface between the 5G RAN and 5GC as NG, and further specified into NG-C and NG-U for CP and UP separately [32]-[34]. NG-C maps to the reference point N2 and NG-U to the reference point N3 [35]. Specifically, as shown in Fig. 4, N2 marks the interface between a gNB and the Access and Mobility Management Function (AMF), and N3 marks the interface between a gNB and the User Plane Function (UPF). In order to support the system architecture alternative 2 described in [36], where our proposed broadcast and multicast user plane network function (XUF) is directly connected to the RAN, a new UP interface M1-NG is introduced, marking the interface between the broadcast and multicast supporting gNB and the XUF. M1-NG is optional and is needed only for system architecture alternative 2.

F1 interface [37] is defined between CU and DU. CUs are interconnected through Xn interface [38]. In 5G UP and CP are clearly separated, and consequently F1 is separated into F1-C on CP and F1-U on UP, Xn into Xn-C on CP and Xn-U on UP. A gNB-CU is further separated logically into gNB-CU-CP on CP, gNB-CU-UP on UP, and gNB-CU-MC connected with the interface E1 [39], where gNB-CU-MC is introduced to support broadcast and multicast, and a new interface F1-M is introduced to connect it with the gNB-DU. In order to support wireless relay by Integrated Access and Backhaul [40], the NG-RAN Un interfaces on both CP and UP are introduced as Un-C and Un-U, to connect the Integrated Access and Backhaul (IAB) nodes. The interfaces within the reference architecture are shown as in Fig. 4. We also introduce Uu as the air interface between the 5G RAN and the UE, to support broadcast and multicast, as shown in Fig. 4.

According to the logical architecture, N2 (NG-C), Xn-C, F1-C and E1 are interfaces on Control Plane. The network interfaces on CP share the same signalling transfer protocol stack, as shown in Fig. 5(a). The Transport Network Layer (TNL) is built on IP [41], [42] transport. Stream Control Transmission Protocol (SCTP) [43] is used for the transport of the application layer signalling protocol.

Interfaces N3 (NG-U), M1-NG, Xn-U, F1-U and F1-M are on User Plane. The network interfaces on UP share the same GTP-U tunnelling protocol stack, as shown in Fig. 5(b). The TNL is built on IP transport and IP Multicast is used for point-to-multipoint delivery of user packets. GTP-U [44] upon User Datagram Protocol (UDP) [45] provides non-guaranteed delivery of UP Protocol Data Unit (PDU) between the gNB and the UPF. N3 fully supports the functions of the M1 interface in LTE, in the cases of 5G-Xcast architecture Alternatives 1 and 3 [14], where broadcast and multicast UP data will be carried over N3 between gNB and UPF. On top of TNL, unicast, multicast and broadcast UP PDUs are multiplexed at Radio Network Layer (RNL).

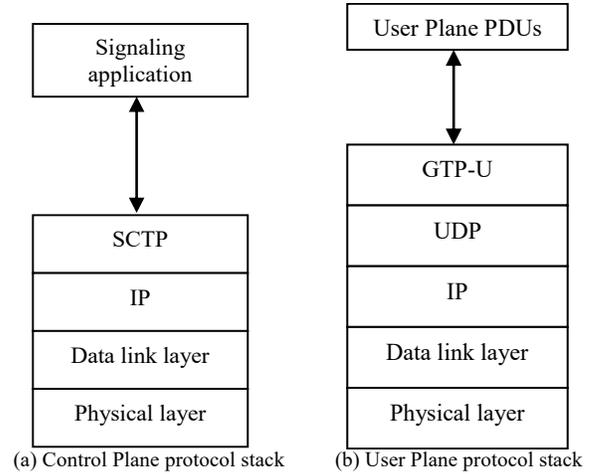

Fig. 5. Interface protocol stack.

## IV. 5G RAN ARCHITECTURE DEPLOYMENTS AND PROCEDURES FOR BROADCAST AND MULTICAST

As PTM services and vertical segments set a variety of very diverse requirements, the design of RAN protocol architecture and procedures should consider the design principles where the multi-service RAN architecture needs to be flexible and support the coexistence of PTP, Single-Cell PTM (SC-PTM), Multi-Cell PTM (MC-PTM) and broadcast transmissions. Baseline for the RAN logical architecture design is NG-RAN Rel-15 architecture.

To allow deployment of existing PTM services and new services, the overall RAN architecture and procedure need to support both (i) dynamic adjustment of the Multicast/Broadcast area based on the user distribution or service requirements and (ii) allow static and dynamic resource allocation between unicast and Multicast/Broadcast. Further, the RAN architecture deployment should support full allocation of downlink carrier resources for Multicast/Broadcast in large geographical areas up to the size of an entire country in SFN mode.

### A. Procedures

The design target of RBMA is to enable dynamic areas based on user geographical distribution, reusing the flexibility of the unicast architecture and basic principles of SC-PTM extended over to MC-PTM. Users having active unicast traffic is in RRC_CONNECTED state [32] and since the UE location is known by a single cell, it is proposed that the RAN (e.g. anchor gNB) should decide the multicast bearer configuration or deliver the multicast traffic over unicast data radio bearers. As shown in Fig. 6, if the number of active users is low, the multicast traffic is delivered to UEs using unicast. When the unicast traffic of a UE is detected to have low activity, the UE is moved to RRC_INACTIVE [46] and the UE continues to receive multicast traffic within the configured RBMA. The RBMA, where the UE can receive multicast traffic, is defined and controlled by RAN and can be part of the Radio Resource Control (RRC) configuration, or part of the broadcasted multicast configuration (e.g. System Information). The anchor gNB (usually the last serving gNB) defines the RBMA configuration, and in case of multiple gNBs, distributes it over Xn interface to the gNBs which belong to the RBMA. Depending on the number of low activity UEs receiving multicast in a cell, the gNB can decide to keep one (or more) UEs in RRC_CONNECTED state, assuming that multicast bearer mapped to unicast bearer or direct usage of unicast bearer is more spectral efficient than multicast bearer in RRC_INACTIVE state with limited feedback. The benefit of RRC_INACTIVE over RRC_IDLE is the maintained connection to AMF / UPF where the connection management state remains in Connection Management (CM)-Connected and the UE Context is stored in both UE and RAN. This will allow low latency state transition between RRC_CONNECTED and RRC_INACTIVE, see Fig. 6.

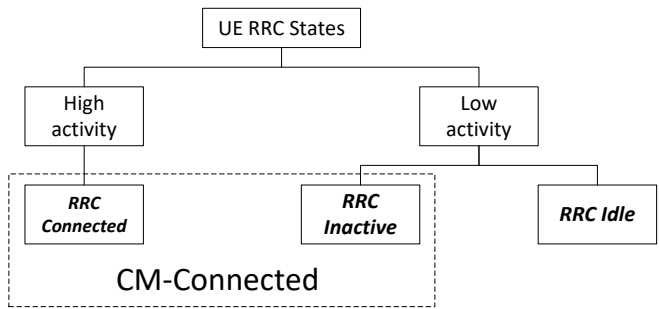

Fig. 6. Unicast activity and RRC States.

An example of the cell selection procedure with two RBMA Ids is presented in Fig. 7, including three UEs receiving IP multicast traffic. This can be further described with three different scenarios:

1. UE1 may be in RRC_CONNECTED state receiving both unicast and PTM multicast traffic from the same DU. Location of UE1 is known by a single cell in RAN, thus enabling the transmission of unicast and multicast traffic using only unicast bearers.
2. The UE2 has completed its unicast traffic, and due to low activity, the RAN (e.g. anchor gNB) decides to suspend the RRC configuration and configures the UE2 into RRC_INACTIVE state. The multicast traffic will move from unicast Radio Bearer (RB) to multicast RB thus allowing the UE2 to continue the reception of PTM multicast traffic. The configuration includes the configuration for RRC_INACTIVE state as well as the PTM Group-Radio Network Temporary Identifier (RNTI) and RBMA Id consisting of at least one cell. The anchor gNB receives the PTM multicast traffic over the N3 data tunnel from UPF. When the UE2 identifies a new cell with better coverage/quality and optionally the current source cell is having degrading coverage/quality, the UE needs to perform a cell reselection to a new cell. As illustrated in Fig. 7, two cases with UE mobility can be identified for UEs in RRC_INACTIVE state.

(a) The UE2 moves inside the RBMA Id1 and performs cell reselection from one DU to another DU. The UE2 does not need notify the network about cell reselection since it is able to receive the same multicast traffic from all transmission points under the same RBMA. The RBMA can consist of one or more gNBs and the UPF traffic is distributed over F1, Xn and N3 interfaces to transmission points to cover the RBMA area. Further, if the RBMA consists of multiple gNB contributing to MC-PTM transmission, then the F1, Xn (necessary synchronization can be controlled by gNB receiving the IP multicast traffic over N3) and N3 interfaces can be used to route the same IP multicast traffic to joining gNBs.

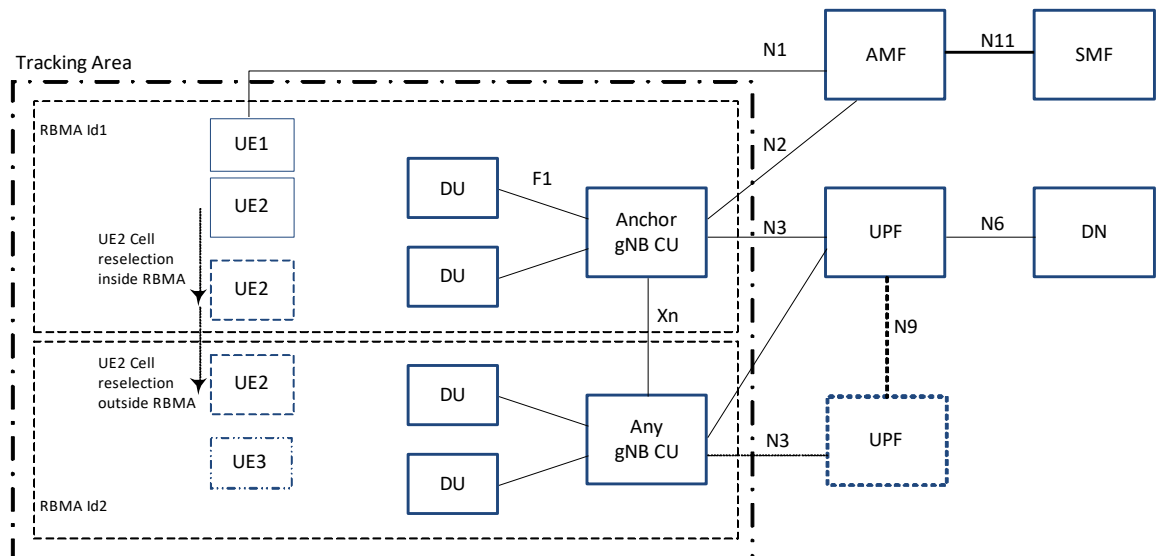

Fig. 7. UE mobility and cell selection/reselection procedure for RAN Broadcast/Multicast Area.

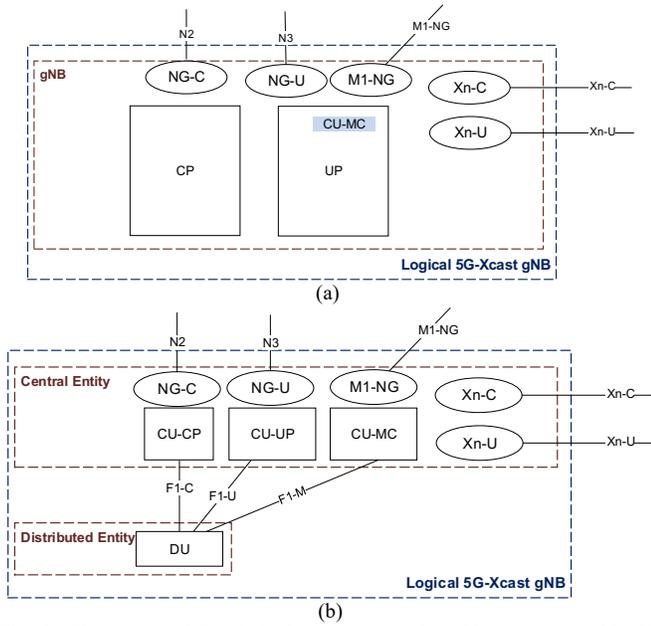

Fig. 8. The proposed RAN deployment scenario without (a) or with (b) functional split.

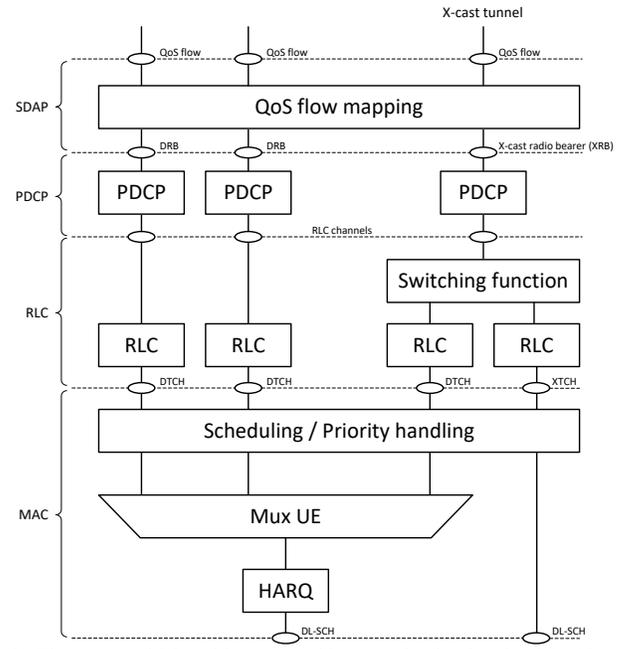

Fig. 9. The proposed L2 architecture and bearer selection in Cloud-RAN.

(b) If the new target cell is outside of the RBMA Id1, UE needs to notify the network its new location with RBMA update. Network will configure the UE with new RBMA Id and if the new RBMA consists of more than one gNB, network performs the RAN based Multicast Area Setup to allow traffic distribution over Xn to gNBs belonging to RBMA.

3. The UE3 is having low unicast activity, its connection towards AMF is released and therefore the UE3 is configured with RRC_IDLE state. In this state the Core Network knows the UE's location only within the tracking area in AMF. Alternatively, the UE3 could be also a receive only device or in Receive Only Mode (ROM) mode without uplink capability, thus the network does not know its existence or location respectively. In these cases, the RBMA may be configured with multiple cells participating in SFN broadcast mode. The RBMA becomes the same as the tracking area or SFN service area and two or more selected cells are participating in SFN, for example according to given pre-configuration. When the area of RBMA Id2 is configured with SFN transmission, all the UEs in that area can benefit from the SFN transmission regardless of their RRC state.

In the case for Terrestrial Broadcast, users are unknown to the RAN (due to the lack of uplink and, therefore, registration into the network) and the RAN can decide beforehand and according to service and coverage requirements the multicast bearer configuration for delivery. From the three scenarios shown above, Terrestrial Broadcast would be an extension of UE3 being it a receive only device with no uplink capability. In this case the RBMA becomes the same as the tracking area or SFN service area. Single-cell transmission of an SFN with multiple cells participating can be configured.

### B. Deployments

Our proposed RAN deployment leverages the major assumptions of 5G NR overall architecture described in [26] which shows RAN architecture for gNBs with and without functional splits.

For the RAN deployment without functional split, all the logical gNB functions as well as RAN interface protocol terminations are hosted in a gNB physical node. Fig. 8(a) depicts this RAN deployment scenario. Herein, the logical nodes include CP and UP. The UP hosts the newly introduced control functions including functions performed by gNB-CU-MC. On the other hand, the UP logical node hosts 5G-Xcast RAN function for delivery of user plane data [19]. The major interface protocol terminations for the aforementioned interfaces are NG-C (to which N2 reference point is mapped), NG-U (to which N3 reference point is mapped), M1-NG, Xn-C and Xn-U.

Fig. 8(b) demonstrates our proposed RAN deployment scenario with functional split. Herein, the figure shows logical nodes (CU-CP, CU-UP and DU), internal to a logical gNB. The major interface protocol terminations for 5G-Xcast interfaces, NG-C, NG-U, M1-NG, Xn-C and Xn-U, are hosted in the central entity. The DU is hosted in a distributed entity. The central entity and distributed entity are separate physical nodes.

In this work, we further propose a Cloud-RAN based deployment. At high level, the DU(s) closer to the deployed cells receive information about a set of UEs to which the multicast data should be transmitted and based on this information the distributed unit configures the needed unicast channels and multicast channels. The CU being a centralised unit and DU a local unit, the DU needs to make the decision of the transmission mode. When the DU receives multicast data from a CU, it will select either unicast or multicast channel to transmit the multicast data to the set of UEs as per the procedures as described in the last subsection.

The proposed Layer 2 radio protocol architecture for Cloud-RAN deployments is shown in Fig. 9. The multicast data is delivered to NG-RAN over a data tunnel, which in this case is referred as X-cast tunnel in Fig. 9 to emphasize the dynamic selection process of RLC entities and transport channels for the transmission. The multicast traffic can comprise of multiple QoS flows. In this case the SDAP can map the QoS flows to a set of newly introduced broadcast/multicast data Radio Bearers (XRB) to enable differentiation at lower layers for different QoS requirements.

The PDCP, which is not used in eMBMS architecture, may provide sequence numbering and duplication detection. In case the UE is receiving the same data over unicast and multicast Dedicated Radio Bearers DRBs, the duplication detection should be supported. The duplication can be used in the proposed architecture also for performance enhancement when the UE receives the same PDCP PDU over Dedicated Traffic Channel (DTCH) and Multicast logical Channel (XTCH) as a means for improving packet reliability. In this case the ciphering functionality used for unicast is not required for multicast. Another PDCP function relevant to the transport of multicast data is the header compression and decompression.

Switching function in the DU is the new functionality proposed to the architecture, where the DU selects the transmission method. Switching function locates below PDCP but above RLC layer, thus not placed in the same Cloud-RAN computing hardware pool as the CU. Thus, using the F1 fronthaul interface, it is natural to place the Switching function in the DU. For a set of UE's receiving multicast data (i.e. the UE's which have expressed their interest in receiving multicast and the PDU session has been modified to allocate the RAN resources for UEs joining in the IP multicast group), a pair of RLC entities and logical channels (i.e., DTCH and XTCH channels) is set up to transmit the multicast data over the air. The multicast logical channels are shared between some or all multicast UEs.

The switching between unicast and multicast can be based on the availability of UE measurements and the reported quantities in the measurements, such as Synchronization Signal - Reference Signal Received Power (SS-RSRP), Channel State Information (CSI)-RSRP, Synchronization Signal - Reference Signal Received Quality (SS-RSRQ), CSI-RSRQ, according to procedures related to RBMA. In general, if measurements are not available, XRB switching is routing traffic though multicast transport channels and when measurement reports indicate poor radio condition for some UEs in comparison to others, the Switching function will select the unicast transport channel for those UEs and the multicast transport channel for other UEs. When setting up an XRB, RRC may configure thresholds in the XRB switching function to select between unicast and multicast logical channels, also considering the minimum number of UEs required for switching to multicast transport and the resulting estimated resource and spectrum efficiency gain.

The gNB-DU switching function configuration includes RLC channels and logical channels for XRB bearer and DL tunnel information. The gNB-DU configures at least unicast transport by creating an RLC entity mapped to a single RLC channel towards PDCP and mapped to a corresponding logical channel in MAC according to DRB setup procedures.

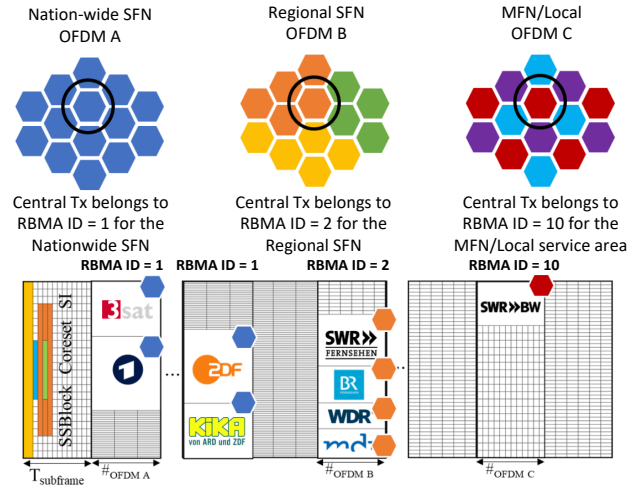

Fig. 10. Three deployments consisting of a nation-wide SFN, a regional SFN and a single cell transmitter and their association to RBMA ID for Terrestrial Broadcast Services.

A new RLC entity and corresponding mapping to a XTCH is created if multicast transport is not configured already. The configuration includes at least one of the following: logical channel identities, RLC configuration (e.g. mode, sequence number field length, timer values), MAC configuration and PHY configuration.

Some examples are provided regarding the cell arrangements from which a broadcast service may be transmitted, in this case assuming TV/radio services. In Fig. 10, three different deployments are shown consisting of a nation-wide SFN, a regional SFN and a deployment covering the same area by means of single cell transmitters. A central hexagon is highlighted, which belongs to different RBMA IDs according to the network planning requirements of each TV/radio service. A frame transmitted from the central hexagon is shown, where, for simplicity, TDM is used to multiplex frames containing the services per different RBMA. The three scenarios are:

- A set of transmitters configured within the same SFN area. In this case a complete carrier (or frame within a carrier) is available to schedule Terrestrial Broadcast services.
- A set of transmitters that constitute different SFN areas requiring synchronization and orthogonal scheduling between SFN areas.
- A set of single-cell transmitters requiring orthogonal scheduling of resources to avoid mutual interference (in this case on a reuse 3 basis).

### C. RAN Network Slicing

One of the key features for the deployment of the proposed RAN architecture in 5G is network slicing. By harnessing network function virtualisation (NFV) and network softwarisation, RBMA can be sliced to facilitate the desired 5G network management solution. As pointed out in [15], it is not appropriate to define a pure multicast slice, as multicast is frequently mixed and tightly integrated with unicast to transport broadcast and multicast communication services. Furthermore, there is a requirement [9] to allow the deployment of a multicast solution that can seamlessly adapt between unicast and multicast transmission to maximise the efficiency of using radio and network resources. However, there is a need to define network slices for a category of broadcast and multicast services on the demand of Communications Service Provider (CSP) and according to the SLA signed with the Network Operator (NOP), or specifically Mobile Network Operator (MNO) for 5G networks.

The RBMA slicing provides a framework to implement the network slicing in 5G-Xcast RAN and sets the ground for future practical deployment as a primary option to provision and manage broadcast and multicast services.

5G RBMA network slicing is the exact solution to meet the requirement specified in 3GPP on 5G MBMS, to *support Multicast/Broadcast network sharing between multiple participating MNOs, including the case of a dedicated MBMS network* [9].

## V. 5G RAN Architecture Evaluation for Terrestrial Broadcast and Multicast

### A. Comparison table

Table I describes the main features of each state-of-the-art cellular broadcast technologies and on-going Rel-17 work, comparing it to the proposed solution.

### B. Imprint analysis

Modified eMBMS SYNC controlled by RAN is proposed to reside between gNB-CU-MC and DU allowing controllable fronthaul latencies. The number of new interfaces impacts directly the service integration and deployment complexity of the new broadcast/multicast system. Possibility to reuse and enhance current NG-RAN interfaces to support broadcast and multicast will keep the complexity low. 3GPP has defined interfaces between the NG-RAN and 5GC and specified reference point N2 for Control Plane and reference point N3 for User Plane. The NG-RAN internal interfaces are those between 5G RAN logical network nodes. Enabling the gNB-CU-C to control the 5G broadcast/multicast and modified gNB-CU-MC as part of the gNB-DU internal interfaces minimises the need for new interfaces [20].

### C. Radio Resource Efficiency

Broadcast/multicast through the 5G Physical Downlink Shared Channel (PDSCH) with basic limited uplink feedback channel allows dynamic deployment of SFN network. SFN transmission involving multiple cells for group transmission improves the spectral efficiency especially at the cell edges when the control of the SFN resides at the gNB-CU-C.

### D. Scalability

Broadcast/multicast with SFN transmission requires one resource allocation for the UE group. In case the SFN service areas are semi-static and no uplink channel feedback is expected from the UEs, the amount of radio resources would be independent of the number of UEs. When the SFN areas are operated in a dynamic manner taking the UE interest in receiving the broadcast/multicast, then resource allocation done per UE group and the dynamic radio resource utilization in SFN is not proportional to the number of users even if the unlimited number of users may not be supported. SFN transmission in NG-RAN is natively supported feature and the SFN broadcast/multicast architecture is integrated into the baseline unicast architecture maximizing the scalability and enabling dynamic switching between different transmission modes for transparent 5G broadcast networks [20].

### E. Dimensionality analysis

The system proposed is formed by one gNB for the entire deployment. Nation-wide SFNs for broadcast are characterised for having a large number of cells, both deployed in High Power High Tower (HPHT) and some used as gap-fillers. The architecture follows a tree-like topology, where one gNB-CU with a gNB-CU-MC serves a large amount of gNB-DU over F1 interface, and the gNB-DUs serve a large number of RRH/cells. In [14], it is specified that the maximum number of uniquely identified gNB-DUs under one gNB-CU allowed by the signalling is $2^{36}-1$, and the maximum number of cells that can be served by one gNB-DU is 512 or $2^9$. Overall, the maximum number of cells served is $(2^{36} - 1)*512$. To the best of authors' knowledge, this value greatly exceeds any existing Digital Terrestrial Television (DTT) deployment [20].

On another vein, the biggest limiting factor for nation-wide SFN deployment in 5G is the Inter-Site Distance (ISD) allowed by New Radio numerologies. As shown in [17], maximum ISD in Rel-15 is 1.41 Km. New physical layer schemes such as the negative numerologies proposed in 5G-Xcast [17] could extend this up to 120 Km, perfectly fit for nation-wide SFN.

### F. Latency analysis

Latency performance parameters in cellular networks are usually divided into Control Plane latency and User Plane latency. In detail, Control Plane latency is the time needed from an idle terminal to switch into a state ready to transmit and/or receive, with enabled context information in RAN and Core Network, while the User Plane latency is the time spent by a packet from the source until it is decoded by the device. Given that one of the design decisions of this architecture was to minimise the imprint over existing 5G solution, the results obtained by 3GPP can be applied to this approach. For standard devices, this Control Plane and User Plane latency can be the same as Rel-15 latency i.e. around 15 ms [11] and 2 ms [17] for Control and User Plane respectively. Possible upgrades to these values can be the use of the newly introduced 5G RRC_INACTIVE state which can lower the overall "wake-up" latency from power-efficient state to active mode, and the use of Multi-access Edge Computing (MEC) to bring the source content closer to the user [20].

TABLE I
COMPARISON BETWEEN CELLULAR BROADCAST TECHNOLOGIES

|  | Air Interface | Single Frequency Network Mode | Dynamic Service Areas | Receive Only Mode | Synchronization used |
|---|---|---|---|---|---|
| MBSFN | LTE | Yes | No | No | SYNC |
| SC-PTM | LTE | No | No | No | SYNC |
| feMBMS | LTE | Yes | No | Yes | SYNC |
| Rel-17 Mixed Mode | NR | No | Yes | No | To be defined |
| 5G-Xcast RAN | NR | Yes | Yes | Yes | RAN-SYNC |

In general, the latency of the proposed architecture design compared to 4G eMBMS can be on the same grade of magnitude since the purpose of SYNC is to compensate for the network deployment delays from geographically away transmitters (e.g. in a nation-wide SFN). In case that the 5G RAN architecture is used to optimise the network resources, it is expected to have a small area SFN with the same latency as 5G unicast.

## VI. CONCLUSION

Having the 5G NR Rel-15 RAN unicast architecture as a basis for our RAN architecture design, we have proposed architectural and functional enhancements allowing a flexible deployment of 5G-Xcast RAN where the new Radio Access Technology (RAT) supports dynamic adjustment of the Multicast/Broadcast geographical area based on e.g. the user distribution or service requirements. The new 5G-Xcast RAN architecture can cover large geographical areas up to the size of an entire country in SFN mode with content synchronization for SFN transmission. Developed RAN Broadcast/Multicast Area and RAN based synchronization solutions can support local, regional and national multicast/broadcast areas. The support for dynamic geographical areas is enabled with the support of not only Terrestrial Broadcast service but also a concurrent delivery of both unicast and multicast/broadcast services to the users, as well as support for efficient multiplexing with unicast transmissions via seamless data bearer selection.

The proposed 5G PTM RAN architecture has been shown to fulfil the 5G-Xcast use case specific requirements [23] and cover the generic architectural requirements listed in 3GPP TS 38.913 [9], as compared in Table II.

Leveraging the proposed solutions in this work can lead to further investigations on generalised RAN framework designs, including more simulations and testing to evaluate the appropriate architecture design for PTM in a more practical scenario. For example, this work suggests that the latency of the proposed solutions can be comparable to that of the 5G unicast, thus one can carry out quantitative evaluations on the potential latency reduction achieved by applying RAN-SYNC or RRC_INACTIVE-assisted wake-up procedure.

TABLE II
5G-XCAST RAN ARCHITECTURE AGAINST 5G BROADCAST/MULTICAST REQUIREMENTS

| Broadcast/Multicast Requirement in 38.913 [9] (RAN architecture related) | 5G-Xcast RAN Architecture solution |
|---|---|
| The new RAT shall support existing Multicast/Broadcast services (e.g. download, streaming, group communication, TV, etc.) and new services (e.g. V2X, etc). | 1. The overall 5G-Xcast RAN architecture supporting multicast/broadcast and vertical segments. 2. PDU Session Modification for multicast resource allocation and multicast bearer selection. |
| The new RAT shall support dynamic adjustment of the Multicast/Broadcast area based on e.g. the user distribution or service requirements. | 1. RBMA to allow dynamic multicast area along with user distribution, requested service, UE unicast activity, mobility and connectivity. 2. RBMA procedures to create and modify RBMA. |
| The new RAT shall support static and dynamic resource allocation between Multicast/Broadcast and unicast; the new RAT shall in particular allow support of up to 100% of downlink resources for Multicast/Broadcast (100% meaning a dedicated MBMS carrier). | 1. RAN based content synchronization supporting Transparent Multicast and Terrestrial Broadcast. 2. Support for SFN with RBMA and RAN based synchronization using gNB-CU-MC enabling synchronised transmission within the gNB-DUs. 3. Multicast deployment using Cloud-RAN for seamless unicast/multicast bearer selection |
| The new RAT shall support Multicast/Broadcast network sharing between multiple participating MNOs, including the case of a dedicated MBMS network. | 1. Support for SFN with RBMA and RAN based synchronization. 2. RAN slicing using RBMA |
| The new RAT shall make it possible to cover large geographical areas up to the size of an entire country in SFN mode with network synchronization and shall allow cell radii of up to 100 km if required to facilitate that objective. It shall also support local, regional and national broadcast areas. | 1. Synchronised multicast content transmission using gNB-CU-MC which enforces the synchronised over the air transmission of multicast/broadcast traffic within the gNB-DUs. 2. Support for SFN with RAN based synchronization |
| The new RAT shall support Multicast/Broadcast services for fixed, portable and mobile UEs. Mobility up to 250 km/h shall be supported. | 1. The overall 5G-Xcast RAN architecture. 2. Support for ROM and PDU Session Modification for multicast resource allocation and multicast bearer selection. |
| The new RAT shall leverage usage of RAN equipment (hard- and software) including e.g. multi-antenna capabilities (e.g. MIMO) to improve Multicast/Broadcast capacity and reliability. | The overall 5G-Xcast RAN architecture, e.g. based on the design principle of maximizing the architectural commonality with unicast. |
| The new RAT shall support Multicast/Broadcast services for mMTC devices. | The overall 5G-Xcast RAN architecture is designed to support all kind of devices, e.g. devices targeting for IP multicast services with TX/RX capability and ROM mode devices. |

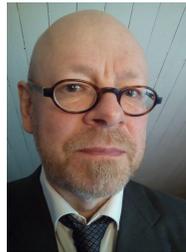
**Mikko Säily** is a 5G research scientist and research project manager leading 5G connectivity, mobility, and positioning research at Nokia Bell Labs, Espoo, Finland. His current research interests include 5G evolution for high accuracy and low-latency positioning and mobility and connectivity for 5G New Radio in millimeter-wave communications. He is the coauthor and coinventor of more than 150 international publications, patents, patent applications, and technical reports.

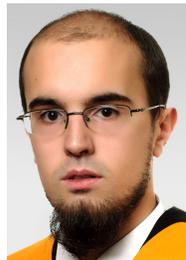
**Carlos Barjau** is a R&D engineer at Mobile Communications Group (MCG) of the Institute of Telecommunications and Multimedia Applications (iTEAM) at Universitat Politecnica de Valencia (UPV). He received a M.Sc. degree in Telecommunications engineering alongside a second M.Sc. degree in Communications and Development of Mobile Services from UPV, Spain, both in 2013. He is working at the moment on his Ph.D. degree on enabling broadcast services in 5G networks and was involved in H2020 project 5G-Xcast. His current research interests include synchronization of radio networks, service-based architectures for broadcast, convergence of terrestrial broadcast and cellular networks and Cloud-RAN deployments.

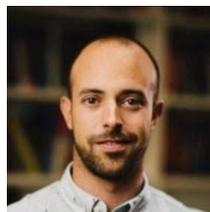
**Jordi J. Gimenez** obtained a Ph.D. in Telecommunications from the Universitat Politècnica de València (UPV) in Spain, while he was a Research Engineer at the iTEAM Research Institute. In 2018 he joined the Institut für Rundfunktechnik (IRT) as Project Manager and Research Engineer for 5G-related projects in the domain of media distribution and contribution, being the H2020 5G-Xcast and 5G-Solutions the most recent. He has been actively contributing to the 3GPP RAN working groups for the standardization of LTE/5G Broadcast and participating in different project groups of the EBU Strategic Programme on Distribution such as 5G-Deployments, which he chairs. Dr. Gimenez has a wide experience on terrestrial broadcast technologies, in particular on physical layer aspects and network planning. He has contributed to several DVB and ATSC technical groups on next-generation terrestrial broadcast technologies such as time-frequency slicing, channel bonding, LDM or WiB.

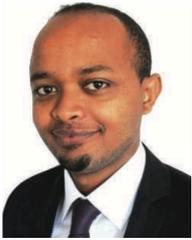

**Fasil Tesema** is a senior research engineer and simulation expert at Nomor Research GmbH, based in Munich, Germany. His research interests are 4G and 5G mobile networks with a focus on radio protocols and resource management for unicast, multicast, and broadcast. He is a Member of the IEEE.

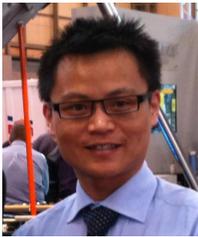

**Wei Guo** received the B.Eng. degree in computer communications and the Ph.D. degree in information and communication systems from the Beijing University of Posts and Telecommunications, China, in 1997 and 2005, respectively. Since 2005, he joined a number of EU funded ICT projects under FP6, FP7, and Horizon 2020. He is currently with the Samsung R&D Institute U.K., where he is involved in several EU funded projects. His research interests include communication protocols, telecommunication network architectures, and deep learning application to telecommunication networks.

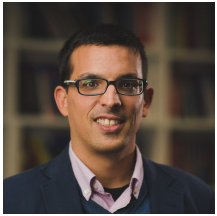

**David Gomez-Barquero** is a Professor with the Communications Department, Universitat Politecnica de Valencia, Spain. He held visiting research appointments with Ericsson Eurolab, Germany, the KTH Royal Institute of Technology, Sweden, the University of Turku, Finland, the Technical University of Braunschweig, Germany, the Fraunhofer Heinrich Hertz Institute, Germany, the Sergio Arboleda University of Bogota, Colombia, the New Jersey Institute of Technology, USA, and the Electronics and Telecommunications Research Institute, South Korea. He participated in digital broadcasting standardization, including DVB-T2, T2-Lite, DVB-NGH and, more recently, ATSC 3.0, as the Vice Chairman of the Modulation and Coding Ad-Hoc Group. He is the Coordinator of the 5G-PPP Project 5G-Xcast, that is, developing broadcast and multicast point-to-multipoint capabilities for the standalone 5G New Radio and the 5G service-enabled Core Network. His current main research interest is the design, optimization, and performance evaluation of next generation wireless communication technologies, including broadcasting.

He is an Associate Editor of the IEEE TRANSACTIONS ON BROADCASTING. He was the General Chair of 2018 IEEE Symposium on Broadband Multimedia Systems and Broadcasting

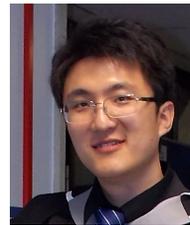

**De Mi** (M'17) received the B.Eng. degree in information engineering from the Beijing Institute of Technology, Beijing, China, in 2011, the M.Sc. degree in communications and signal processing from Imperial College London, U.K., in 2012, and the Ph.D. degree from the 5G Innovation Centre (5GIC), Institute for Communications Systems (ICS), University of Surrey, U.K., in 2017. He is currently a Research Fellow in future wireless communications and Academic Ph.D. Supervisor of Engineering and Physical Sciences Research Council (EPSRC) Industrial Cooperative Awards in Science & Technology (CASE) programme at the University of Surrey. He has been the RAN work package leader of 5G-PPP project 5G-Xcast and Project Coordinator of industrial collaborative project SoftRAN in 5GIC/ICS. His current main research interests include air interface design, multiantenna signal processing, broadcast and multicast technologies, and millimetre-wave communications.